\documentclass[conference,10pt]{IEEEtran}
\IEEEoverridecommandlockouts
\usepackage{cite}
\usepackage{xcolor,colortbl}

\def\BibTeX{{\rm B\kern-.05em{\sc i\kern-.025em b}\kern-.08em
		T\kern-.1667em\lower.7ex\hbox{E}\kern-.125emX}}
\usepackage[utf8]{inputenc}
\usepackage[T1]{fontenc}
\usepackage{amsmath,amssymb,amsfonts}
\usepackage[abbreviations]{foreign}
\usepackage{algorithm}
\usepackage{algpseudocode}

\usepackage{xspace}

\usepackage{listings}
\newcommand{\code}[1]{\texttt{\small #1}\xspace}

\usepackage{subcaption}
\usepackage{textcomp}
\usepackage{booktabs}

\usepackage{url}
\usepackage{pifont}
\usepackage[colorlinks=true,linkcolor=blue,anchorcolor=blue,citecolor=blue,filecolor=black,menucolor=black,runcolor=black,urlcolor=blue]{hyperref}
\usepackage{lipsum}

\usepackage[inline]{enumitem}
\setlist{noitemsep,topsep=0pt,parsep=0pt,partopsep=0pt}

\usepackage[capitalise]{cleveref}

\usepackage{tikz}

\definecolor{lightcolor}{rgb}{0,0.5,1}

\usepackage{fontawesome5}
\newboolean{showcomments}
\setboolean{showcomments}{true}
\ifthenelse{\boolean{showcomments}}
{ \newcommand{\mynote}[3]{
		\fbox{\bfseries\sffamily\scriptsize#1}
		{\small$\blacktriangleright$\textsf{\emph{\color{#3}{#2}}}$\blacktriangleleft$}}}
{ \newcommand{\mynote}[3]{}}

\definecolor{darkgreen}{rgb}{0.3,0.5,0.3}
\definecolor{darkblue}{rgb}{0.3,0.3,0.5}
\definecolor{darkred}{rgb}{0.5,0.3,0.3}

\newcommand{\ocalls}{\texttt{ocall}s\xspace}
\newcommand{\ocall}{\texttt{ocall}\xspace}
\newcommand{\ecalls}{\texttt{ecall}s\xspace}
\newcommand{\ecall}{\texttt{ecall}\xspace}
\newcommand{\tlibc}{\texttt{tlibc}\xspace}
\newcommand{\libc}{\texttt{libc}\xspace}
\newcommand{\rbf}{\texttt{rbf}\xspace}
\newcommand{\rbs}{\texttt{rbs}\xspace}

\newcommand{\fread}{\texttt{fread}\xspace}
\newcommand{\fwrite}{\texttt{fwrite}\xspace}
\newcommand{\fopen}{\texttt{fopen}\xspace}
\newcommand{\fclose}{\texttt{fclose}\xspace}
\newcommand{\sread}{\texttt{read}\xspace}
\newcommand{\swrite}{\texttt{write}\xspace}
\newcommand{\iread}{\texttt{i-read}\xspace}
\newcommand{\iwrite}{\texttt{i-write}\xspace}
\newcommand{\fseeko}{\texttt{fseeko}\xspace}
\newcommand{\ifread}{\texttt{i-fread}\xspace}

\newcommand{\ifwrite}{\texttt{i-fwrite}\xspace}
\newcommand{\ifseeko}{\texttt{i-fseeko}\xspace}

\newcommand{\ifrw}{\texttt{i-frw}\xspace}
\newcommand{\ifrwoc}{\texttt{i-frwoc}\xspace}
\newcommand{\ifoc}{\texttt{i-foc}\xspace}
\newcommand{\ifr}{\texttt{i-fr}\xspace}
\newcommand{\ifw}{\texttt{i-fw}\xspace}

\newcommand{\iall}{\texttt{i-all}\xspace}

\newcommand{\nosl}{\texttt{no\_sl}\xspace}
\newcommand{\zc}{\texttt{zc}\xspace}

\newcommand{\memcpy}{\texttt{memcpy}\xspace}
\newcommand{\vanilla}{\texttt{vanilla-memcpy}\xspace}
\newcommand{\zcmemcpy}{\texttt{zc-memcpy}\xspace}
\newcommand{\lmbench}{\texttt{lmbench}\xspace}
\newcommand{\kissdb}{\texttt{kissdb}\xspace}
\newcommand{\openssl}{\texttt{OpenSSL}\xspace}

\newcommand{\addfigvspace}{\vspace{-3mm}}

\lstdefinestyle{CStyle}{
	backgroundcolor=\color{backgroundColour},   
	commentstyle=\color{mGreen},
	keywordstyle=\color{magenta},
	numberstyle=\tiny\color{mGray},
	stringstyle=\color{mPurple},
	basicstyle=\footnotesize,
	breakatwhitespace=false,         
	breaklines=true,                 
	captionpos=b,                    
	keepspaces=true,                 
	numbers=left,                    
	numbersep=5pt,                  
	showspaces=false,                
	showstringspaces=false,
	showtabs=false,                  
	tabsize=2,
	language=C
}

\newcounter{numobserv} 
\definecolor{beaublue}{rgb}{0.88, 0.93, 0.93}
\usepackage{tcolorbox}
\colorlet{shadecolor}{beaublue}
\tcbset{
	colback=beaublue,
	boxrule=0.5pt,
	boxsep=0pt,
	left=1pt,
	right=1pt,
	top=1pt,
	bottom=1pt,
	sharp corners,
	colframe=white,
}
\newcommand{\observ}[1]{
	\addtocounter{numobserv}{1}
	\begin{tcolorbox}	
		\textit{\textbf{Take-away\,\thenumobserv\,:} #1 }	
	\end{tcolorbox}
}

\newcommand{\sys}{\textsc{ZC-Switchless}\xspace}

\newcommand{\copyrighttext}{  \scriptsize \textcopyright 2023 IEEE.               
	Personal use of this material is permitted.                                 
	Permission from IEEE must be obtained for all other uses,                   
	in any current or future media, including reprinting/republishing this      
	material for advertising or promotional purposes, creating new collective   
	works, for resale or redistribution to servers or                           
	lists, or reuse of any copyrighted component of this work in other works.   
	Pre-print version. Published in the 53rd Annual IEEE/IFIP International Conference on Dependable Systems and Networks (DSN).
	
}

\begin{document}
	
	\title{SGX Switchless Calls Made Configless}

	\author{\IEEEauthorblockN{Peterson Yuhala}
		\IEEEauthorblockA{\textit{University of Neuchâtel}\\
			Neuchâtel, Switzerland \\
			peterson.yuhala@unine.ch}
		\and
		\IEEEauthorblockN{Michael Paper}
		\IEEEauthorblockA{\textit{ENS de Lyon} \\
			Lyon, France \\
			michael.paper@ens-lyon.fr}
		\and
		\IEEEauthorblockN{Timothée Zerbib}
		\IEEEauthorblockA{\textit{Institut Polytechnique de Paris} \\
			Paris, France \\
			timothee.zerbib@ip-paris.fr}
		\and 
		
		\IEEEauthorblockN{Pascal Felber}
		\IEEEauthorblockA{\textit{University of Neuchâtel} \\
			Neuchâtel, Switzerland \\
			pascal.felber@unine.ch}\\
		\and	
		
		\IEEEauthorblockN{\hspace{3cm}}
		\IEEEauthorblockA{~\\
			~\\
		}
		\and
		\IEEEauthorblockN{Valerio Schiavoni}
		\IEEEauthorblockA{\textit{University of Neuchâtel} \\
			Neuchâtel, Switzerland \\
			valerio.schiavoni@unine.ch}
		\and		
		\IEEEauthorblockN{Alain Tchana}
		\IEEEauthorblockA{\textit{Grenoble INP} \\
			Grenoble, France\\
			alain.tchana@grenoble-inp.fr}	
		\and	
		
		\IEEEauthorblockN{\hspace{1cm}}
		\IEEEauthorblockA{~\\
			~\\
		}
	}

\newcommand{\copyrightnotice}{\begin{tikzpicture}[remember picture,overlay]       
	\node[anchor=south,yshift=2pt,fill=yellow!20] at (current page.south) {\fbox{\parbox{\dimexpr\textwidth-\fboxsep-\fboxrule\relax}{\copyrighttext}}};
	\end{tikzpicture}
}
	
	\maketitle
	\copyrightnotice
	\thispagestyle{plain}
	\pagestyle{plain}

\begin{abstract} 
	Intel's software guard extensions (SGX) provide hardware enclaves to guarantee confidentiality and integrity for sensitive code and data. 
	However, systems leveraging such security mechanisms must often pay high performance overheads.
	A major source of this overhead is SGX enclave transitions which induce expensive cross-enclave context switches.
	The Intel SGX SDK mitigates this with a \emph{switchless} call mechanism for transitionless cross-enclave calls using worker threads. 
	Intel's SGX switchless call implementation improves performance but provides limited flexibility: developers need to statically fix the system configuration at build time, which is error-prone and misconfigurations lead to performance degradations and waste of CPU resources.
	\sys is a configless and efficient technique to drive the execution of SGX switchless calls. 
	Its dynamic approach optimises the total switchless worker threads at runtime to minimise CPU waste.
	The experimental evaluation shows that \sys obviates the performance penalty of misconfigured switchless systems while minimising CPU waste.
\end{abstract}

\section{Introduction}
\label{sec:introduction}


Cloud computing increases privacy concerns for applications offloaded to the cloud, as sensitive user data and code are largely exposed to potentially malicious cloud services.
Trusted execution environments (TEEs) like Intel SGX~\cite{innoTech,innovInst13,hoekstra2013using,vcostan} provide secure \emph{enclaves}, offering confidentiality and integrity guarantees to sensitive code and data. 
Enclaves cannot be accessed by privileged or compromised software stacks, including the operating system (OS) or hypervisor. 
However, SGX enclaves introduce overheads, primarily from \emph{enclave context switches}, upon every CPU transition between non-enclave and enclave code. 
The Intel SGX SDK~\cite{sgxsdk} provides specialised function call mechanisms (\ie \ecalls and \ocalls), to enter, and respectively exit, an enclave. 
Enclave switches cost up to 14,000 CPU cycles~\cite{hotcalls,sgxperf}, on average 56$\times$ more expensive compared to regular system calls on similar Intel CPUs (\ie 250 cycles~\cite{eleos}).
This represents a serious performance bottleneck for applications which perform many enclave context switches, \eg to perform system calls via \ocalls~\cite{hotcalls}.

Techniques exist~\cite{scone,hotcalls,switchless} which circumvent expensive enclave switches by leveraging \emph{worker threads} in and out of the enclave. 
\emph{Client threads}, \ie inside or outside of the enclave, send their requests to the opposite side via shared memory. 
Worker threads handle such requests and send the appropriate responses once completed. 
The Intel SGX SDK implements this technique via the \emph{switchless call library}~\cite{sgxdevref}.
While this approach improves the performance of enclave context switches, our practical exeperience revealed  the following problems.
First, Intel SGX switchless calls \textbf{must be manually configured} at build time and misconfigurations can lead to performance degradations and waste of CPU resources~\cite{sgxdevref}. 
Second, manually configuring \ecalls or \ocalls as switchless at build time is not ideal: developers rarely know the frequency nor the duration of the calls, which are typically application and workload specific.


To mitigate these problems, we propose \sys (\emph{zero-config} switchless), a system to \textbf{dynamically select switchless routines} at run time and \textbf{configure the most appropriate number of worker threads on the fly}. This approach prevents static configuration of switchless routines and worker threads, and obviates the performance penalty of misconfigured switchless systems, while minimizing CPU waste.

%

\sys leverages an application level scheduler that monitors runtime performance metrics (\ie number of wasted cycles or fallback calls to regular non-switchless routines, \etc); these are used to dynamically configure an optimal number of worker threads to fit the current workload while minimising CPU waste.

We further analysed the Intel SGX SDK implementation, and derived practical configuration tips for Intel SGX switchless calls. In addition, we tracked performance issues with Intel's SDK \texttt{memcpy} implementation, and propose an optimised version which removes a major source of performance overhead for both switchless and regular SGX transition routines.


In summary, our \textbf{contributions} are: 
\begin{enumerate*}[noitemsep,nolistsep,leftmargin=*,label=(\arabic*)]
\item The design and the implementation of \sys, a configless and efficient system to dynamically select and configure switchless calls, to be released as open-source.
\item An optimised implementation for the Intel SGX SDK's \texttt{memcpy}, also to be open-sourced. 
\item An extensive experimental evaluation demonstrating the effectiveness of our approach via micro- and macro-benchmarks.
\end{enumerate*}


\section{Background on Intel SGX}
\label{sec:background}


Intel SGX extends Intel's ISA to enable secure memory regions called \emph{enclaves}.
The CPU reserves an encrypted portion of DRAM, called the \emph{enclave page cache} (EPC), to store enclave code and data at runtime.
EPC pages are only decrypted inside the CPU by the \emph{memory encryption engine} (MEE) once loaded in a CPU cache line. 

Enclaves operate only in user mode and thus cannot issue system calls (syscalls)~\cite{vcostan} directly. 
The \ecalls and \ocalls are specialised function calls to enter (using the \code{EENTER} CPU instruction) and exit (using the \code{EEXIT} CPU instruction) the enclave, respectively.
The Intel SGX SDK provides a secure version of the C standard library, the \emph{trusted libc} (\tlibc), for in-enclave applications. 
The \tlibc only relies on trusted functions, and removes instructions forbidden by Intel SGX~\cite{sgxdevref}. 

Enclave applications can be deployed following two approaches:
\emph{(1)}~running the entire application in the enclave (\eg SCONE \cite{scone}, Graphene-SGX~\cite{grapheneSGX}, Occlum~\cite{occlum}, SGX-LKL~\cite{sgxlkl}), or
\emph{(2)}~splitting the application into a trusted and untrusted part which respectively execute inside and outside of the enclave (\eg Glamdring~\cite{glamdring}, Civet~\cite{tsai2020civet}, Montsalvat~\cite{yuhala2021montsalvat}). 

For both cases, unsupported routines not implemented by the \tlibc must be relayed to the untrusted part via \ocalls, to be executed in non-enclave mode.

Both \ecalls and \ocalls perform expensive CPU context switches, \ie switching from enclave to non-enclave mode, and vice-versa. 
The overhead of enclave transitions is mainly due to the CPU flushing its caches as well as all TLB entries containing enclave addresses, so as to preserve confidentiality~\cite{vcostan}. More precisely: an \ocall $=$ \code{EEXIT} $+$ \code{untrusted host processing} $+$ \code{EENTER}. \code{EEXIT} flushes the CPU caches and possibly invalidates branch predictors and TLBs. Similarly, the \code{EENTER} instruction performs many checks and requires hardware-internal synchronization of CPU cores~\cite{graphene-docs}. 

In the following, we focus on \ocalls as they are usually the principal source of overhead due to enclave context switches, based on our experience.  
However, the techniques we propose in this paper can equally be used for \ecalls.

\begin{figure}[!t]
	\centering
	\includegraphics[scale=0.55]{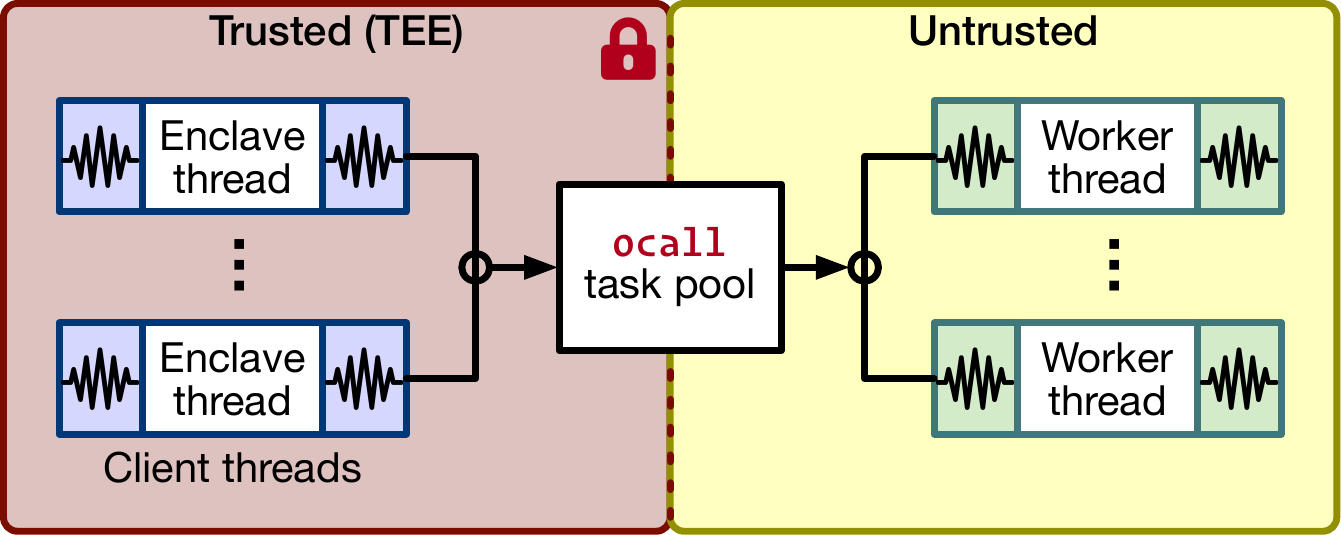}
	\caption{Intel SGX switchless \ocall architecture.}
	\label{fig:ocall-arch}
\end{figure}

\smallskip\noindent
\textbf{Switchless ocalls.}
The Intel SGX SDK provides two variants of \ocalls.
First, \emph{regular ocalls} perform costly context switches (up to 14,000 cycles~\cite{sgxperf}) from enclave mode to non-enclave mode. 
Second, \emph{switchless ocalls}~\cite{sgxdevref} provide a mechanism to do cross-enclave calls without performing an enclave context switch. 
In a nutshell (\ie \autoref{fig:ocall-arch}), \emph{worker threads} outside the enclave perform unsupported functionality (\eg syscalls) on behalf of in-enclave client threads. 
The latter send \ocall requests to a task pool in untrusted memory. 
The worker threads outside the enclave wait for pending tasks in the task pool, and execute them until the task pool is empty~\cite{sgxdevref}. 
These operations are done without performing any enclave transition. 

\label{ocalls}

\section{Limitations of Intel SGX switchless calls}
\label{sec:motivations}

In the following, we identify three main problems and limitations with the Intel switchless calls implementation:
\emph{(i)}~switchless call selection (\S\ref{intel-switchless-call-selection}),
\emph{(ii)}~worker thread pool sizing (\S\ref{worker-thread-pool-sizing}), and
\emph{(iii)}~Intel SDK parameterisation (\S\ref{intel-sdk-parametrization}).

\smallskip\noindent
\textbf{Setup.} We use a 4-core (8 hyper-threads) Intel Xeon CPU E3-1275 v6 clocked at 3.8\,GHz, 8\,MB L3 cache, 16\,GB of RAM, with support for Intel SGX v1. 
We deploy Linux Ubuntu 18.04.5 LTS with kernel 4.15.0-151 and Intel SGX SDK v2.14.
We report the median over 10 executions.

\subsection{Switchless calls selection}
\label{intel-switchless-call-selection}

At build-time, developers must specify the routines (\ie \ecalls or \ocalls) to be handled switchlessly at run time.
The Intel SGX reference~\cite{sgxdevref} suggests to configure a routine as switchless if it has \emph{short} duration and is \emph{frequently called}.
However, these details are hardly available to the developer at build time. 
Auto-tuning tools~\cite{sgxTuner} cannot spot potential switchless routines by relying solely on duration and frequency, as it would require a full code path exploration, a costly operation for large systems.
Further, the execution frequency of a specific application routine is workload specific and hard to configure at build time.

Improperly selected switchless routines degrade the performance of SGX applications.
We show this behaviour with a synthetic benchmark.
It executes $n$ \ocalls to 2 functions as follows:
$\alpha$ calls to function $f$ which is known to benefit from switchless calls (as shown in \cite{switchless}), while $\beta$ calls to $g$ which should run as a regular \ocall.
If $\alpha = \beta$, the sum of the execution times of calls to $f$ is negligible compared to the same sum for calls to $g$.
Hence, to better highlight the performance gains when executing $f$ switchlessly, we set $\alpha = 3\beta$ and $n = \alpha + \beta$. 
In this test, $f$ is an empty function (\ie, \texttt{void f(void)\{\}}).
On the other hand, $g$ routine executes \texttt{asm("pause")} in a loop, \ie a busy-wait loop.

We evaluate five different configurations: \texttt{C1}-\texttt{C5}. 
In \texttt{C1}, all $f$ functions run switchlessly, while $g$ functions run as regular \ocalls.
We expect \texttt{C1} to perform best.
In \texttt{C2}, only the $g$ functions run switchlessly, and we expect \texttt{C2} to be the worst. 
In the \texttt{C3} case, $\frac{\alpha}{2}$ $f$ and $\frac{\beta}{2}$ $g$ functions run switchlessly while the other $f$ and $g$ functions run as regular \ocalls.
Finally, in \texttt{C4} and \texttt{C5} all functions run switchlessly or regularly, respectively.
We observe the following results when executing 100'000 \ocalls (\ie switchless and regular \ocalls combined).
\texttt{C1} is the fastest configuration (0.9\,s), $\approx$ 1.8$\times$ faster than \texttt{C2}, indeed the worst (1.6\,s).
\texttt{C3} and \texttt{C4} complete in 1.3\,s (1.44$\times$ slower than \texttt{C1}).
Finally, \texttt{C5} completes in 1\,s.

\observ{An improper selection of switchless coroutines degrades the performance of SGX applications.}

\begin{figure}[!t]
  \includegraphics[scale=0.7]{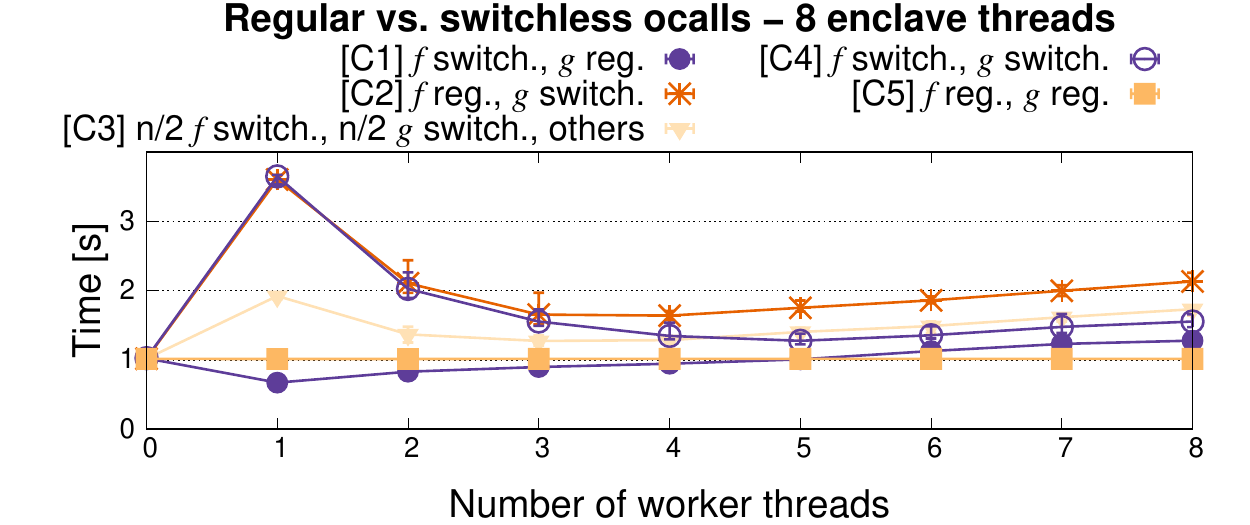}
  \caption{Runtime for 75,000 switchless \ocalls to $f$ and 25,000 regular \ocalls to $g$}
  \label{fig:worker-thread-pool-sizing-assessement-results}
  \vspace{-3mm}
\end{figure}

\subsection{Worker thread pool sizing}
\label{worker-thread-pool-sizing}

Developers must specify the total number of worker threads allowed to the SGX switchless call mechanism at build time. 
However, an overestimation of worker threads for \ocalls can lead to a waste of CPU resources due to busy-waiting of worker threads awaiting switchless requests (see~\cite{sgxdevref}, page 71).
The latter will limit the number of applications that can be co-located on the same server or interfere with application threads which will be deprived of CPU resources.
Similarly, an underestimation can lead to poor application performance as more \ocalls will perform costly enclave switches.

Using the same synthetic benchmark from \S\ref{intel-switchless-call-selection}, we validate these effects for a varying number of worker threads (see \autoref{fig:worker-thread-pool-sizing-assessement-results}).
The optimal number of workers strictly depends on the specific configuration.
At its best (\texttt{C1}), the fewer the workers, the better the performance: this is expected, as there are as many in-enclave threads as hardware cores.
In the other cases, we observe better results when executing functions as regular \ocalls, where the best overall result is using 4 worker threads.

\begin{figure}[!t]
  \includegraphics[scale=0.7,trim={0 40pt 0 0}]{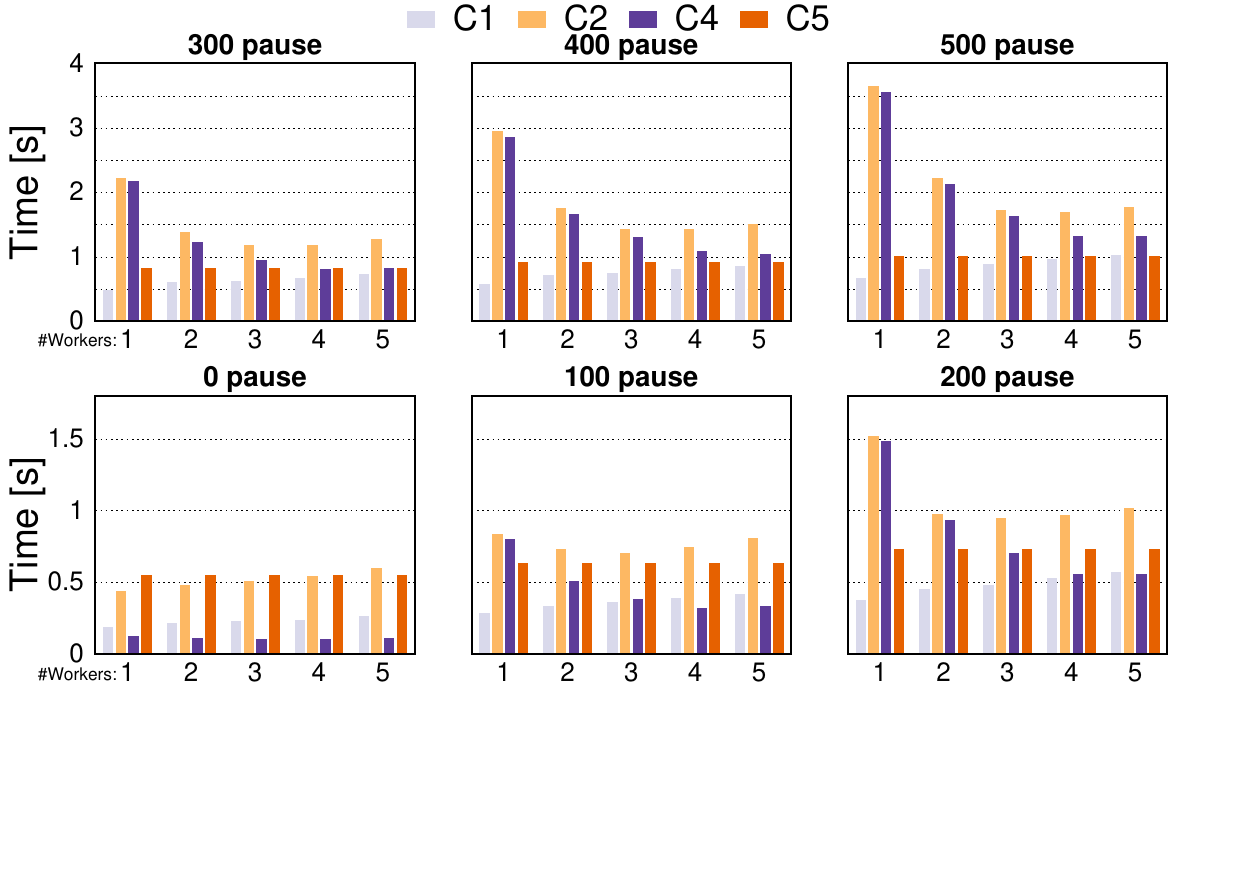}
 \caption{
	Runtime for $100,000$ \ocalls with 8 in-enclave threads for different durations of $g$ function.
}
  \label{fig:worker-thread-pool-sizing-assessement-results-extended}
  \addfigvspace
\end{figure}

\autoref{fig:worker-thread-pool-sizing-assessement-results-extended} presents the execution time of the application depending on both the number of worker threads (from 1 to 5) and the duration of $g$ (from 0 to 500 \texttt{asm\{"pause"\}} instructions, each spending 140 CPU cycles).
We report 4 configurations (we omit \texttt{C3} for the sake of clarity and because it obtains results between the worst of \texttt{C1} and \texttt{C2}).
We can make the following observations.
\texttt{C5} (\ie, both $f$ and $g$ as regular \ocalls) performs worst on average for the shortest $g$ function (\ie 0 pauses), but it is best in several cases for longer $g$ functions, regardless of the number of workers.
\texttt{C1} ($f$ switchless, $g$ regular) is the best when $g$ is longer than 200 pauses.
Executing all functions switchlessly (\texttt{C4}) is good for short $g$ functions, scaling with the available worker threads.

\observ{Switchless calls perform best when the calls are short, relative to the cost of an enclave transition.}

%

\subsection{Intel SDK parameterisation}
\label{intel-sdk-parametrization}

If a switchless task pool is full or all worker threads are busy, a switchless call falls back to a regular \ecall or \ocall.
The Intel SGX SDK defines the variable \texttt{retries\_before\_fallback} (\rbf) for the number of retries client threads perform in a busy-wait loop, waiting for a worker thread to start executing a switchless call request, before falling back to a regular \ecall/\ocall~\cite{sgxdevref}. 
Similarly, the SDK defines \texttt{retries\_before\_sleep} (\rbs), for the number of \texttt{asm\{"pause"\}} done by a worker while waiting for a switchless request, before going to sleep. 
The SDK's default values for both \rbf and \rbs are set at 20,000 retries.

However the value of \rbf especially is abnormal in both a theoretical (as we explain next) and practical sense:
between successive retries, a caller thread executes an \texttt{asm\{"pause"\}} instruction, which has an estimated latency up to 140 cycles on Skylake-based microarchitectures (where SGX extensions were first introduced, see ~\url{https://intel.ly/3hTVEMG}, page 58).
A caller thread can hence wait more than 2.8\,M cycles before its call is handled by a worker thread. 
This is about 200$\times$ more costly relative to a regular \ocall transition ($\approx$14,000 cycles), and defeats the purpose of using switchless calls, \ie to avoid the expensive transition.
Similarly for \rbs, a worker thread will wait for 2.8\,M cycles before going to sleep.

While developers can easily tune the \rbf and \rbs values at build time, the Intel SDK lacks proper guidance.

\observ{The proper configuration of the Intel SGX switchless-related parameters remains hard and misconfigurations lead to poor performance.}

\section{\sys}
\label{sec:contributions}

The main goal of \sys is to provide a resource-efficient implementation for SGX switchless calls.

\noindent
We first present its internal scheduler in \S\ref{scheduler}-\ref{switchless-call-selection}.
In addition, we detail a more efficient implementation of \tlibc's \texttt{memcpy} (\S\ref{memcpy-optimization}), used for intra-enclave data copying, as well as data exchange between the enclave and the outside world.


\begin{figure}
  \centering
  \includegraphics[scale=0.55]{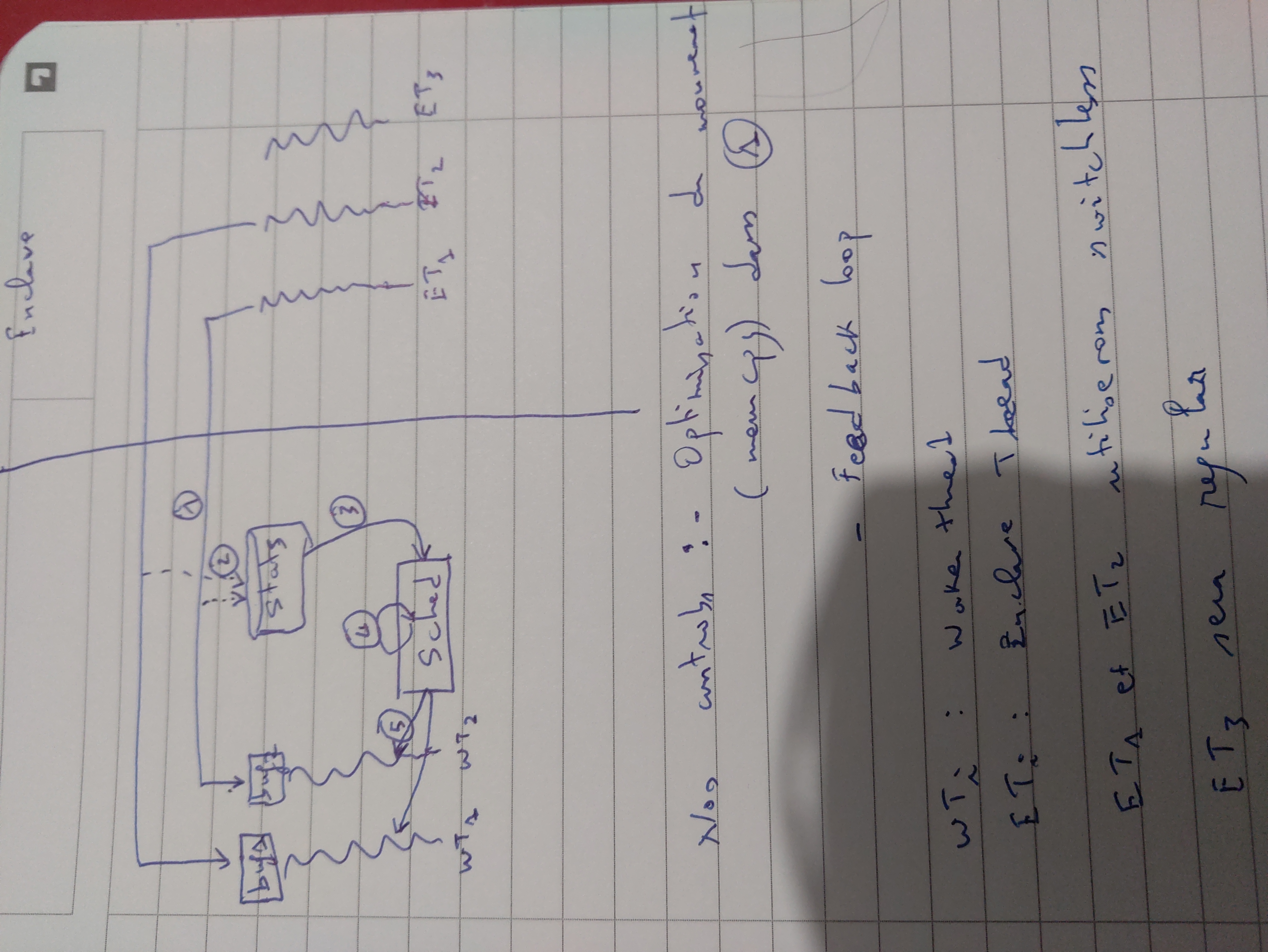}
  \caption{\sys general overview.}
  \label{fig:architecture}
\end{figure}

\autoref{fig:architecture} shows an overview of \sys.
In a nutshell, we consider \textit{any function} (\autoref{fig:architecture}-\ding{182}) as a potential candidate to run as switchless, thus avoiding the need for manual selection by developers at build time.
Our design allows the number of worker threads to be tuned dynamically, to minimise CPU waste while improving performance via the switchless call technique.
We do so by having the scheduler implement a feedback loop (\autoref{fig:architecture}-\ding{183}) to periodically collect \ocall statistics (\autoref{fig:architecture}-\ding{184}), determine the optimal number of worker threads (\autoref{fig:architecture}-\ding{185}), and finally apply its decision (Figure~\ref{fig:architecture}-\ding{186}).
Inside the enclave, a call is executed in a switchless manner if the caller finds at least one idle worker thread.
The remainder details further these aspects.
Unless indicated otherwise, the term \emph{scheduler} refers to \sys's scheduler.
%
%

\subsection{\sys's scheduler}
\label{scheduler}

The main objective of the scheduler is to minimise wasted CPU cycles. 
We define a \emph{wasted CPU cycle} as one spent by a CPU core doing something that does not make the application (\ie caller thread) move forward in its execution~\cite{10.1145/2901318.2901326}. 

In the case of Intel SGX, we identify two potential sources of wasted CPU cycles:
\emph{(1)}~transitions between enclave and non-enclave mode in the case of regular \ocalls, and
\emph{(2)}~busy-waiting in the case of switchless \ocalls.
The overhead of regular \ocalls has been evaluated extensively in past research~\cite{hotcalls}, and it varies also according to the specific CPU and micro-code version.
We evaluated this overhead to be $\sim$13,500 CPU cycles for our experimental setup (see \S\ref{sec:motivations}).

Each active worker can be at any point in time in one of two states:
either the worker is handling an \ocall, in which case the enclave thread (which made the \ocall) is busy-waiting,
or the worker is busy-waiting for incoming \ocall requests.
Therefore, for every active worker thread, there is always exactly one thread busy-waiting.
The extra cost of having $M$ worker threads is thus $M$ multiplied by the number of cycles during which they have been active.

\begin{figure}[!t]
	\centering
	\includegraphics[scale=0.55]{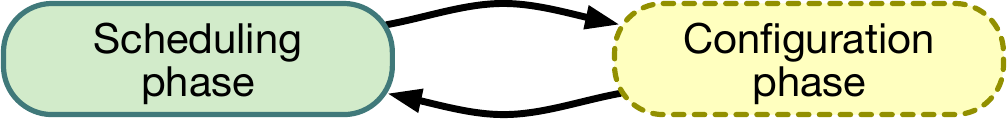}
	\caption{\sys scheduler phases.}
	\label{fig:sched-phases}
\end{figure}

The scheduler periodically computes the number of worker threads that minimises the number of wasted CPU cycles. 
Throughout its lifetime, the scheduler switches between two phases (see \autoref{fig:sched-phases}): a \emph{scheduling phase} that lasts a scheduler quantum, $Q$, during which it sets an optimal number of switchless workers, and a \emph{configuration} phase, during which it calculates the optimal number of switchless workers for the next scheduling phase.

We denote by $T_{es}$ the duration of an enclave switch, $F$ the number of calls not being handled switchlessly (\ie fallback), $N$ the number of cores on the machine and $M$ the number of worker threads set during \sys's scheduler quantum ($Q$, set empirically to 10\,ms). 

The number of wasted cycles during $T$ cycles is: $U = F \cdot T_{es} + M \cdot T$.
At every quantum (\ie during a \emph{configuration phase}), the scheduler thread estimates the optimal number of workers for the next quantum (\ie \emph{scheduling phase}).

To estimate the number of workers during a \emph{configuration phase}, the scheduler sleeps for $\frac{N}{2}+1$ micro-quanta, of length $\mu \cdot Q$ each, with a different number of workers $i$ each time, $0 \leq i \leq \frac{N}{2}$ (\ie $\frac{N}{2}+1$ possible values). The constant $\mu$ is a small time period, so the configuration phase can be quick, but still long enough to capture the needs (in terms of CPU resources) of the application at a given time. We empirically set $\mu = \frac{1}{100}$.
During every micro-quantum, the number of non-switchless calls is recorded so that the scheduler can compute $U_i = F_i \cdot T_{es} \,+\, i \cdot \mu \cdot Q \cdot \textit{CPU\_FREQ}$ once awake.
Finally, the scheduler keeps $M'$ workers for the next \emph{scheduling phase}, where $M'$ is such that $U_{M'} = \min_i U_i$. 



To deactivate a worker thread, the scheduler sets a value in the worker's buffer (see \S\ref{wt-state-machine}).
The worker's loop function will eventually check this value and, if it is set and no caller thread has reserved (or is using) the worker, the worker will pause.
To re-activate a paused worker, the scheduler sends a signal to wake up the corresponding worker thread.


\begin{figure}[!t]
  \centering
  \includegraphics[scale=0.55]{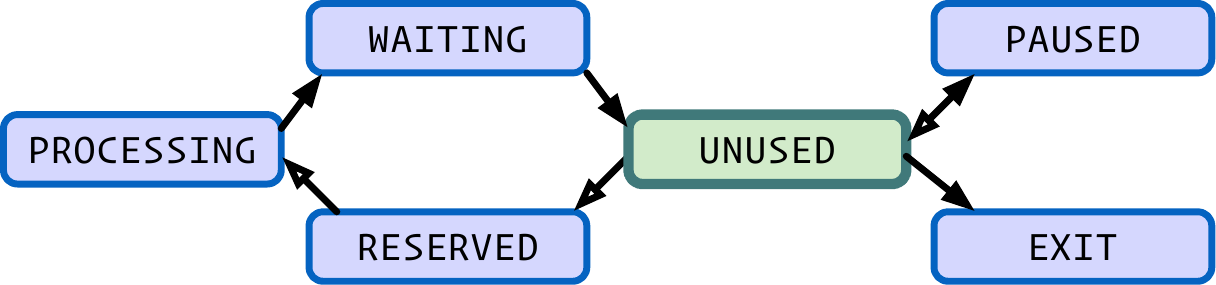}
  \caption{Worker thread state transitions.}
  \label{fig:status_transitions}
  \addfigvspace
\end{figure}

\subsection{Worker thread state machine}
\label{wt-state-machine}

We associate to each worker a \texttt{buffer} structure that consists of 4 main fields: an untrusted memory pool (preallocated) used by callers to allocate switchless requests, a field to hold the most recent switchless request, a status field to track the worker status, and a field used to communicate with the scheduler.
\autoref{fig:status_transitions} summarises the \texttt{status} transitions of a worker thread.

A worker is initially in the \texttt{UNUSED} state.
When a caller needs to make a switchless call, it finds an \texttt{UNUSED} worker and switches the worker's state to \texttt{RESERVED}.
We rely on GCC atomic built-in operations~\cite{atomics} for thread synchronisation. 
The caller allocates a switchless request structure from the corresponding memory pool.
A switchless request comprises: an identifier of the function to be called, the function arguments (if present), and the return result (if present). 

The caller copies its request to the worker's buffer and changes the worker's state from \texttt{RESERVED} to \texttt{PROCESSING}. 
At this point, the worker reads the request and calls the desired function with the corresponding arguments. 
Once the function call completes, the worker updates the request with the returned results (if present) and switches from the \texttt{PROCESSING} to the \texttt{WAITING} state.

Finally, the caller copies the returned results into enclave memory and changes the worker's state to \texttt{UNUSED}.

The memory pools of worker buffers are freed and re-allocated when full via an ocall.
Using preallocated memory pools prevents callers from performing \ocalls to allocate untrusted memory for each switchless request, which will defeat the purpose of using a switchless system. 

Upon program termination, the scheduler sets a value in workers' buffers so the workers can switch to the \texttt{EXIT} state. 
At this stage, the workers perform final cleanup operations (\eg freeing memory) and then terminate.

\subsection{Switchless call selection}
\label{switchless-call-selection}
In \zc, any routine can be run as switchless if the corresponding enclave caller thread finds an available/unused worker thread. Otherwise, the call immediately falls back to a regular \ocall without any busy waiting.

\subsection{Integrating ZC switchless with other TEE implementations}
Other popular TEE implementations operate following a very similar architecture to Intel SGX. For example in ARM TrustZone (Armv8-M architecture)~\cite{demystifying-tz}, the application is divided into two parts: a secure world (\ie enclave) with very limited system functionality, and a normal world (untrusted world) with a richer system API. Similar to Intel SGX, CPU thread transitions between the secure and normal worlds require extra security checks to guarantee data confidentiality and integrity. So conceptually, the design proposed in ZC can be applied here. This will, however, require some modifications at the implementation level.

\subsection{Security analysis of \sys}
The security analysis of \sys is similar to that presented in \cite{hotcalls}. All switchless call designs (\ie Intel switchless, \sys, hotcalls~\cite{hotcalls}) are based on threads in and out of the enclave communicating via plaintext shared memory. Thus, the switchless design proposed by \sys is no less secure than that proposed by the Intel SGX switchless library. 

\smallskip
\textbf{Impact of ZC scheduler on security.} The \sys scheduler is located in the untrusted runtime, which leaves it vulnerable to malicious tampering. However, since the scheduler only decides how many worker threads should be used and when, the worst case scenario here will be a DoS, \eg by killing worker threads. The enclave's confidentiality or integrity, however, cannot be compromised by tampering with the scheduler.

	


\begin{figure}[!t]
  \includegraphics[scale=0.7]{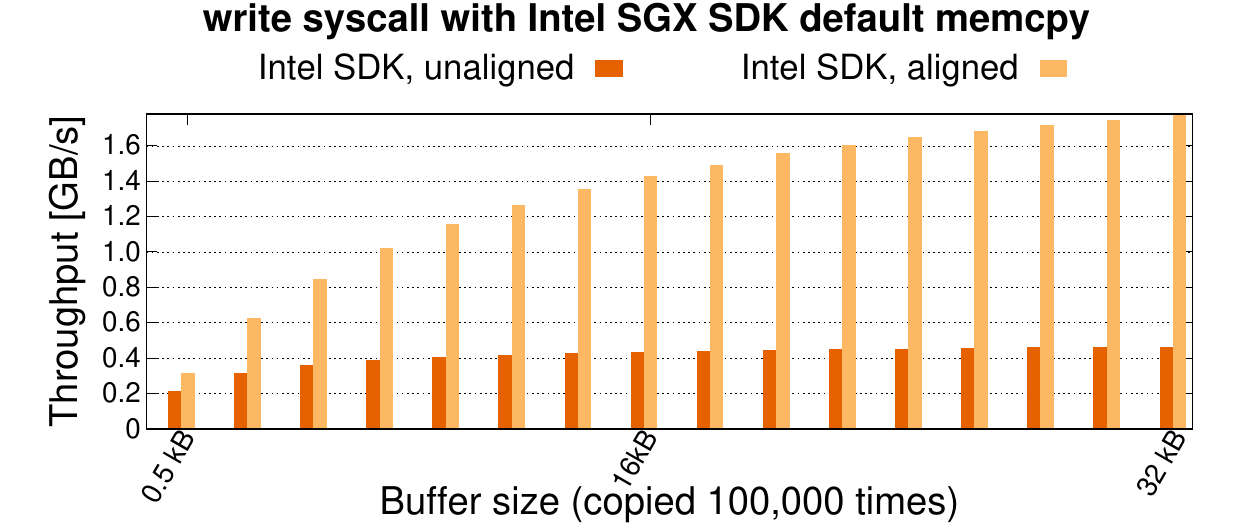}
  \caption{Throughput for \ocalls of the \texttt{write} system call to \texttt{/dev/null} (100,000 operations, average over 10 executions) for aligned and unaligned buffers.} 
  \label{fig:write-bench}
  \vspace{-2mm}
\end{figure}

\subsection{Trusted libc's \texttt{memcpy} optimisation}
\label{memcpy-optimization}

For security reasons, the Intel SDK provides its own implementation of a subset of the functions from the \libc, \ie the \tlibc.
The \tlibc re-implements most of the functions from the \libc that do not require system calls, \eg \texttt{memset}, \texttt{memcpy}, \texttt{snprintf}, \etc.
Because enclave code cannot be linked to dynamic libraries, these re-implementations are statically linked to the enclave application at build time.

We focus on the Intel SDK \tlibc version of \texttt{memcpy}~\cite{sgxmemcpy}, as it is heavily used to pass \ocall arguments from trusted memory to untrusted memory and back for the results~\cite{haibo}.
Our tests highlighted huge performance gaps when using aligned buffers (\ie when the \texttt{src} and \texttt{dest} arguments are congruent modulo 8) and unaligned buffers.

For instance, \autoref{fig:write-bench} presents the throughput when issuing 100,000 \texttt{write} system calls, which requires an \ocall and involves \texttt{memcpy} calls, with varying lengths of aligned and unaligned buffers ranging from 512\,B to 32\,kB.
We observe that the execution time for unaligned buffers is consistently higher than for aligned buffers.
Moreover, when using unaligned buffers, we observe poor scalability trends for \texttt{write} when increasing the buffer sizes, basically plateauing at about 0.4\,GB/s.

By analysing Intel's original implementation of \tlibc's \texttt{memcpy}, we observed that it performs a \emph{software word-by-word copy for aligned buffers} but \emph{a byte-by-byte copy for unaligned buffers}.
We provide a revised and more efficient implementation leveraging the hardware copy instruction \texttt{rep movsb}, as also advised by Intel's optimization manual~\cite{sgxmanual}. 
\autoref{lst:zc-memcpy} sketches our optimised approach, which we evaluate in depth in~\S\ref{eval:memcpy}.

\begin{figure}[t!]
\begin{lstlisting}[language=C,label={lst:zc-memcpy},
caption=\sys optimised \texttt{memcpy}.]
void *memcpy(void *dst0, const void *src0, size_t length){
    ...
    /* Copy forward. */(*@\\[-2pt]@*)
    (*@\color{lightcolor}\_\_asm\_\_ volatile(@*)
        (*@\color{lightcolor}"rep movsb"@*)
        (*@\color{lightcolor}: "=D"(dst0), "=S"(src0), "=c"(length)@*)
        (*@\color{lightcolor}: "0"(dst0), "1"(src0), "2"(length)@*)
        (*@\color{lightcolor}: "memory");@*)(*@\\[-2pt]@*)
done:
    return (dst0);
}
\end{lstlisting}
\addfigvspace
\end{figure}

\section{Evaluation}
\label{sec:evaluation}

Our evaluation answers the following questions:

\begin{itemize}[]
	\item[$Q_1$)] How does \sys impact application performance for (1) static workloads (\S\ref{eval:zc-perf-static}), and (2) dynamic workloads (\S\ref{eval:zc-perf-dynamic})?		
	\item[$Q_2$)] What is the effect of misconfigurations of Intel switchless on application performance? (\S\ref{eval:zc-perf-static}) and (\S\ref{eval:zc-perf-dynamic})?
	\item[$Q_3$)] What is the effect of \sys on CPU utilisation for static (\S\ref{eval:zc-cpu-static}) and dynamic (\S\ref{eval:zc-cpu-dynamic}) workloads? 	
	\item[$Q_4$)] What is the performance gain of our improved \memcpy implementation (\S\ref{eval:memcpy}) ?

\end{itemize}

\smallskip\noindent\textbf{Experimental setup.} All experiments use a server equipped with a 4-core Intel Xeon CPU E3-1275 v6 clocked at 3.8\,GHz with hyperthreading 
enabled.
The CPU supports Intel SGX, and ships with 32\,KB L1i and L1d caches, 256\,KB L2 cache and 8\,MB L3 cache.
The server has 16\,GB of memory and runs Ubuntu 18.04 LTS with Linux kernel version 4.15.0-151.
We run the Intel SGX platform software, SDK, and driver version v2.14.
All our enclaves have maximum heap sizes of 1\,GB.
The EPC size is 128\,MB (93.5\,MB usable by enclaves).
We use both static and dynamic benchmarks. 

Our static benchmarks are based on \texttt{kissdb}~\cite{kissdb} and an Intel SGX port of \texttt{OpenSSL}~\cite{sgxssl}: \texttt{kissdb} is a simple key/value store implemented in plain C without any external dependencies, while \texttt{OpenSSL}~\cite{openssl} is an open-source software library for general-purpose cryptography and secure communication.

For dynamic benchmarks, we use \texttt{lmbench}~\cite{mcvoy1996lmbench}, a suite of simple, portable, ANSI/C microbenchmarks for UNIX/POSIX. 

\smallskip
For all benchmarks, we set the initial number of worker threads to $\frac{\#logical\_cpus}{2}$ for \sys.
This number is 4 for the SGX server used.
For Intel switchless experiments, we maintain the default \rbf and \rbs values (\ie $20,000$).

\subsection{Static benchmark: \kissdb}
We issue a varying number of key/value pair writes to \texttt{kissdb}, and we evaluate and compare the performance and CPU utilisation of \sys with Intel switchless.

We ran our benchmark in 3 modes:
without using switchless calls (\nosl),
using Intel switchless calls, and
using \sys (\zc for short).
For Intel switchless, we consider two values for the number of switchless worker threads: 2 and 4.
In \kissdb, we know empirically that the 3 most frequent \ocalls in the benchmarks are: \fseeko, \fwrite, and \fread.
Therefore, for \kissdb we benchmark Intel's switchless in 10 ($2 \times 5$) different configurations:
only \fseeko as switchless (\ifseeko-$x$, $x$ being 2 or 4, the number of Intel switchless worker threads),
only \fwrite as switchless (\ifwrite-$x$),
only \fread as switchless (\ifread-$x$),
both \fread and \fwrite as switchless (\ifrw-$x$), and
all the 3 \ocalls as switchless (\iall-$x$).
Note that these ten configurations correspond to possible configurations an SGX developer could have set up for \kissdb. We report the averages over $5$ runs.

\subsubsection{\sys \emph{vs.} Intel switchless\label{eval:zc-perf-static}}
\emph{Answer to Q1 \& Q2.} \autoref{fig:kissdb-perf}~(a) and~(b) show the average latencies for setting a varying number of 8-byte keys and 8-byte values in \kissdb, with respectively 2 and 4 switchless worker threads configured for Intel switchless.

\begin{figure}[!t]
	\vspace{-4mm}
	\centering
	\includegraphics[scale=0.6]{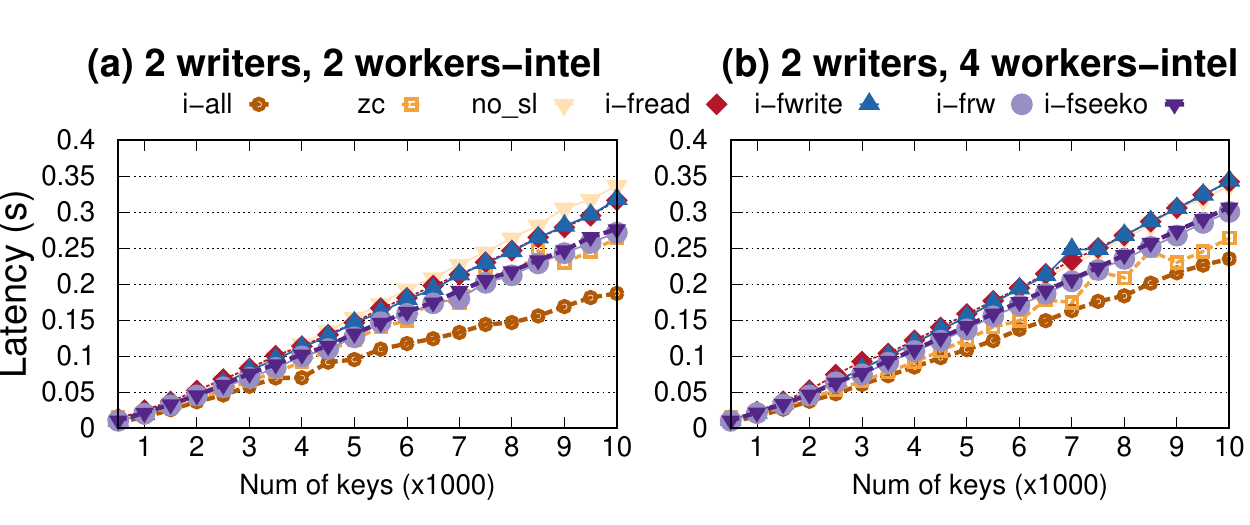}
	\caption{\texttt{kissdb}: average latency of key/value \texttt{SET} commands.}
	\label{fig:kissdb-perf}
	\addfigvspace
\end{figure}

\smallskip\noindent\textbf{Observation.}
Here, \zc is $1.22\times$ faster when compared to a system without switchless calls, and respectively $1.19\times$, $1.13\times$,  $1.13\times$, $1.05\times$, and $1.02\times$ faster when compared to \ifread-2, \ifwrite-2, \ifseeko-2, and \ifrw-2.

However, \zc is about $1.33\times$ slower for \iall-2.
We note how \zc is (on average) $1.26\times$ faster than \ifread-4, $1.22\times$ faster than \ifwrite-4, $1.13\times$ faster than \ifseeko-4, $1.10\times$ than \ifrw-4.
However, it is $1.16\times$ slower than \iall-4.

\smallskip\noindent
\textbf{Discussion.}
In \kissdb, \fseeko is the most frequent \ocall, invoked almost twice more often than \fread and \fwrite. Further, \fseeko is much shorter in duration relative to \fread and \fwrite, which explains the better performance of \ifseeko when compared to \ifread and \ifwrite for both 2 and 4 switchless worker threads.
\ifwrite configurations show the poorest performance for Intel's switchless in all cases.

However, when both \fread and \fwrite are configured as switchless (\ifrw), we see an improvement in performance relative to \ifwrite and \ifread, almost equal to \ifseeko performance. Here the combined sum of \fread and \fwrite calls surpasses the number of \fseeko invocations, which leads to a more significant number of switchless calls in \ifrw, thus leading to similar performance as \fseeko.

This configless strategy of \zc outperforms statically misconfigured systems like \ifread and \ifwrite (for both 2 and 4 Intel switchless workers), shows similar performance as \ifseeko-2, and is faster than \ifseeko-4.
The observed spikes (\eg 7,500 and 8,500 keys) in \zc are due to \ocall operations when reallocating full memory pools for \zc buffers (see \S\ref{wt-state-machine}).

\observ{\sys achieves better performance relative to non-switchless systems, and outperforms misconfigured Intel switchless systems.}

In \iall, Intel's switchless outperforms \zc because it maintains a constant number of switchless workers (2 or 4), whereas \zc uses few worker threads at various points during the application's lifetime.

\observ{A well configured Intel switchless system outperforms \zc, but \zc obviates the performance penalty observed from the misconfigured switchless sytems.}

\subsubsection{\sys vs. Intel switchless: CPU usage}
\label{eval:zc-cpu-static}

\emph{Answer to Q3.}
We now evaluate the CPU utilisation of \sys and Intel switchless when running the same experiments as in \S\ref{eval:zc-perf-static}.
We measured the overall CPU utilisation for the given systems from the kernel's \texttt{/proc/stat}.
The percentage CPU utilisation is calculated by: 
$\%cpu\_used =\frac{ (user+nice+system)}{ (user+nice+system+idle)} * 100$
where:
\emph{user} is the time spent for normal processes executing in user mode,
\emph{nice} is the time spent for processes executing with ``nice'' priority in user mode,
\emph{system} is the time spent for processes executing in kernel mode, and
\emph{idle} is time spent by the CPU executing the \textit{system idle process}.


\autoref{eval:cpu-usage-kiss} shows the average CPU usage for setting a varying number of 8-byte keys and 8-byte values in \kissdb, with respectively 2 and 4 switchless worker threads configured for Intel switchless.

\begin{figure}[!t]
	\centering
	\addfigvspace
	\includegraphics[scale=0.6]{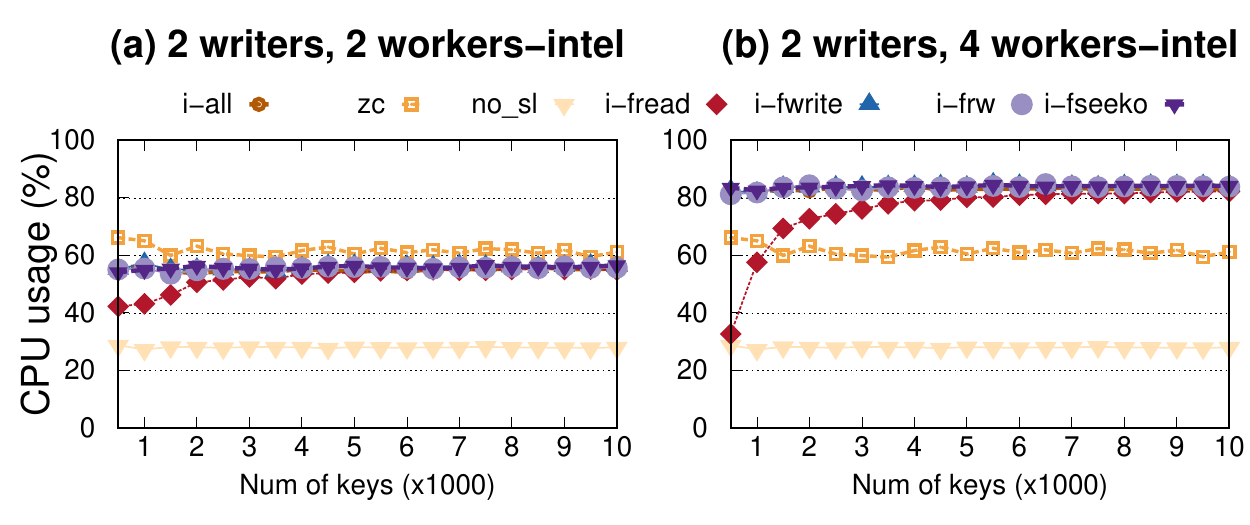}
	\caption{Average \%CPU usage for key/value pair \texttt{SET} ops in \kissdb}
	\label{eval:cpu-usage-kiss}
	\vspace{-1.5mm}
\end{figure}

\smallskip\noindent
\textbf{Observation.}
The experimental results show that \zc maintains approximately $60\%$ CPU usage throughout the benchmark's lifetime.
For 2 Intel switchless workers, all Intel switchless configurations stabilise at about $55\%$ CPU usage, while for 4 Intel switchless workers, the Intel switchless configurations stabilise at about $80\%$ CPU usage.
All switchless systems have visibly higher CPU usage when compared to the system without switchless calls enabled (\nosl).

\smallskip\noindent
\textbf{Discussion.}
Intel's switchless mechanism maintains a constant number of worker threads (2 or 4) throughout the application's lifetime, while \zc's scheduler increases or decreases the number of worker threads (to a maximum of 2 or 4) with respect to the workload. This explains the overall lower CPU usage of \zc relative to Intel switchless.



\observ{\zc outperforms misconfigured systems (\eg \ifread,\ifwrite,\ifrw) while minimizing CPU usage.}

\subsection{Static benchmark: \openssl file encryption/decryption}
This benchmark consists of two enclave threads encrypting and decrypting data read from files. The first thread reads chunks of plaintext from a file, encrypts these in the enclave, and writes the corresponding ciphertext to another file, while the second thread reads ciphertext from a different file, and decrypts it inside the enclave. All cryptographic operations are done with the AES-256-CBC~\cite{aes256} algorithm.

Similarly, we ran this benchmark in the 3 modes described above: \nosl, using Intel switchless calls (2 and 4 workers), and \zc.

In the \openssl benchmark, we know empirically that 4 \ocalls are called most frequently:  \fread, \fwrite, \fopen, and \fclose. We consider 10 possible configurations ($2 \times 5$) for Intel's switchless routines: only \fread as switchless (\ifr-$x$, $x$ being 2 or 4, the number of Intel switchless worker threads), only \fwrite as switchless (\ifw-$x$), both \fread and \fwrite (\ifrw-$x$), both \fopen and \fclose (\ifoc-$x$), and all 4 \ocalls as switchless (\ifrwoc-$x$).

\autoref{eval:2-lat-cpu-openssl} and \autoref{eval:4-lat-cpu-openssl} show the latency and CPU usage for these configurations. We highlight and discuss essential observations.

\begin{figure}[!t]
	\centering	
		
	\begin{subfigure}[b]{0.9\columnwidth}
		\hspace{2mm}
		\includegraphics[scale=0.6]{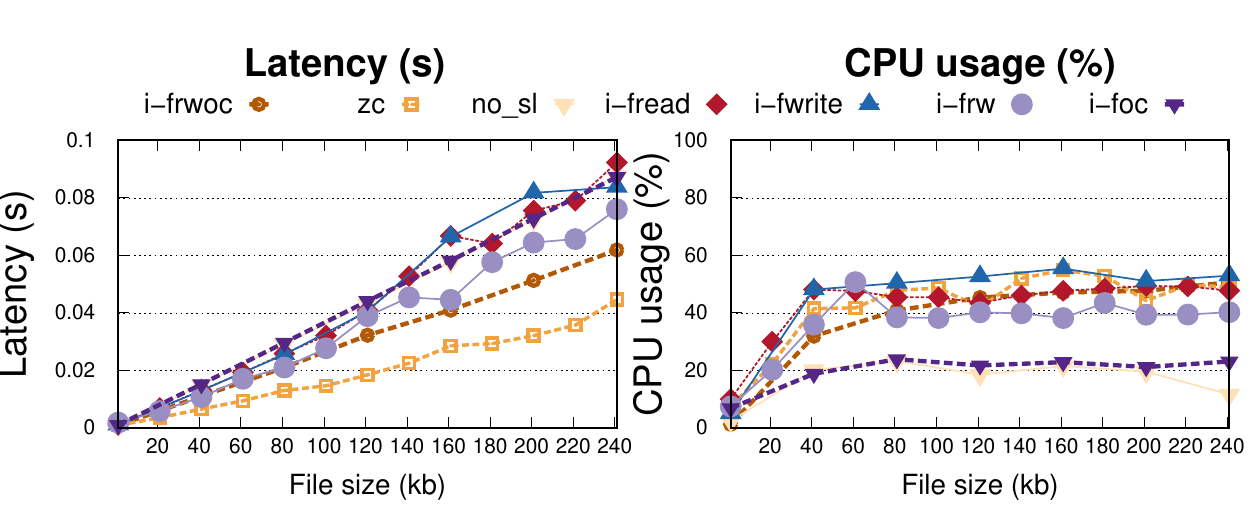}
		\caption{2 Intel workers.}
		\label{eval:2-lat-cpu-openssl}
	\end{subfigure}
	\begin{subfigure}[b]{0.9\columnwidth}
		\hspace{2mm}
		\includegraphics[scale=0.6]{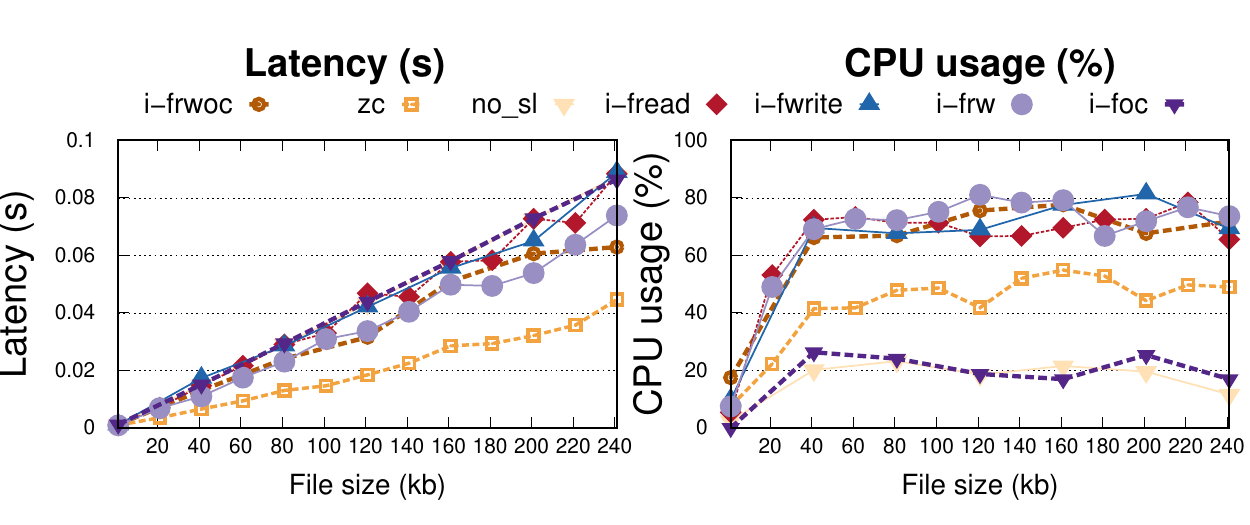}	
		\vspace{-4mm}
		\caption{4 Intel workers.}
		\label{eval:4-lat-cpu-openssl}
	\end{subfigure}
	\vspace{3mm}
	\caption{Latency and CPU usage for \openssl.}
	\label{eval:openssl}
\end{figure}

\smallskip\noindent
\textbf{Observation and discussion.}
In this \openssl benchmark, \fread and \fwrite are respectively called $\approx 2700\times$ and $\approx 1400\times$ more frequently as compared to both \fopen and \fclose. This explains why Intel's latency is poor (close to \nosl) with \ifoc, as more ocalls perform context switches, and much better (w.r.t \nosl) with \ifrw, where a larger number of ocalls are performed switchlessly. Intel performs best with \ifrwoc, when the four ocalls are all configured as switchless. However, we observe that \zc is about $1.62\times$ and $1.82\times$ faster than Intel's best configuration \ifrwoc, for 2 and 4 Intel workers respectively. This is explained by the fact that the \fread and \fwrite calls here are much longer (about $6\times$ longer as compared to the previous \kissdb benchmark). This accentuates the bad effect of the poor default \rbf value in Intel's switchless, as enclave threads do longer pauses waiting for a switchless worker thread to become available. This is absent in \zc, where caller threads immediately fall back.

Regarding CPU usage, \zc's scheduler set the number of worker threads to $0$, $1$, $2$, $3$, $4$ for respectively $9.4\%$, $4.6\%$, $84.4\%$, $1.6\%$, and $0\%$ of the program’s lifetime. This explains the similar CPU usage in \zc and Intel 2 workers (except for \ifoc), while with 4 Intel workers, CPU usage for Intel's best config is about $1.62\times$ larger than \zc's, despite the latter performing better.

\observ{\zc outperforms all Intel configurations when \ocalls are long; this is due to the poor default \rbf value in Intel's switchless library.}

\subsection{Dynamic benchmark: \lmbench}

%

Our dynamic benchmark is based on the \texttt{read} and \texttt{write} system call benchmarks of \lmbench. The \texttt{read} benchmark iteratively reads one word from the \texttt{/dev/zero} device~\cite{mcvoy1996lmbench}, while the \texttt{write} benchmark iteratively writes one word to the \texttt{/dev/null} device. We devised a dynamic workload approach which consists of periodically (every $\tau = 0.5s$) issuing a varied number of read and write operations to \texttt{lmbench} using two in-enclave caller threads (1 reader + 1 writer) over a period of $60s$. These operations trigger \ocalls, and \zc scheduler adapts the number of worker threads accordingly.

The total run time of the dynamic benchmark is divided into 3 distinct phases, each lasting $20s$: \textit{(1) increasing operation-frequency}: the number of operations is doubled  periodically, \textit{(2) constant operation-frequency}: the number of operations remains at a constant value (the peak value from phase-1), and \textit{(3) decreasing operation-frequency}, where the number of operations is periodically decreased (reduced by half every $\tau$). We measure the read/write throughputs and CPU usage at different points during the benchmark's lifetime.

Similarly, we ran our benchmark in 3 modes:
without using switchless calls (\nosl), using Intel switchless calls, and using \sys (\zc).
For Intel switchless, we consider two values for the number of switchless worker threads: 2 and 4. 
For \lmbench syscall benchmark, we know empirically that the \sread and \swrite syscalls are the most frequent \ocalls. So we configure Intel's switchless in six ($3 \times 2$) different configurations:
only \swrite as switchless (\iwrite-$x$),
only \sread as switchless (\iread-$x$),
both \sread and \swrite \ocalls as switchless (\iall-$x$).
Similarly, these six configurations correspond to possible configurations an SGX developer could have set for their program. We show the throughputs and CPU usages as observed from both the reader and writer threads. We highlight essential observations, and analyze them in our discussions.

\begin{figure}[!t]
	\centering	
		\includegraphics[scale=0.6]{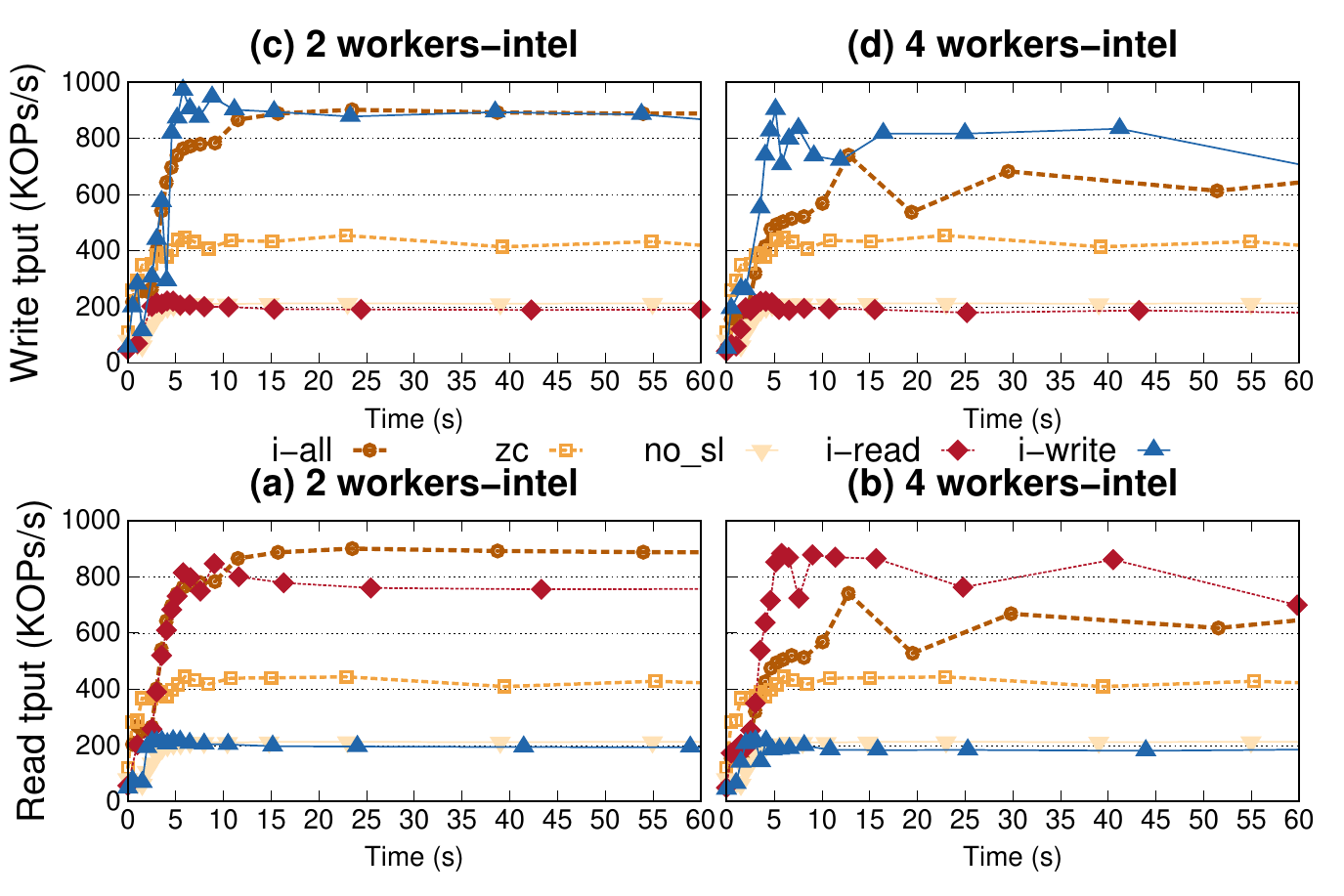}
		\label{eval:tput-lmbench-writer}
	\caption{Read (a/b), write (c/d) tput. for \zc \emph{vs.} 2/4 Intel switchless worker threads.}
	\label{eval:tput-lmbench-reader-writer}
	\addfigvspace
\end{figure}

\subsubsection{Dynamic benchmark: \sys vs. Intel switchless}
\label{eval:zc-perf-dynamic}
\emph{Answer to Q1 \& Q2.}
\autoref{eval:tput-lmbench-reader-writer} shows the operation throughputs as observed by the reader (bottom) and writer (top) threads respectively during the lifetime of the dynamic benchmark. 

\smallskip\noindent

\textbf{Observation.}
The experimental results show that on average, \zc is about $2.3\times$ faster when compared to both reader-\iwrite-2 and reader-\iwrite-4, and $2.1\times$ and $2.5\times$ faster when compared to writer-\iread-2 and writer-\iread-4 respectively.
However, \zc is about $1.6\times$ and $1.1\times$ slower when compared to properly configured Intel switchless configurations \iall-2 (reader,writer) and \iall-4 (reader, writer) respectively.

\textbf{Discussion.}
From the perspective of the reader thread, \iwrite is a misconfiguration as the \texttt{read} calls will never be switchless, and similarly for the writer thread, \iread is a misconfiguration as the \texttt{write} calls will never be switchless. This explains the relatively lower throughputs of these configurations when compared to \zc and the other switchless configurations. However, \zc has a lower throughput when compared to the better configurations: reader-(\iall,\iread) and writer-(\iall,\iwrite).

\subsubsection{\sys CPU utilisation vs. Intel switchless}
\label{eval:zc-cpu-dynamic}

\emph{Answer to Q3.}
We compute the CPU usage as explained previously. 
\autoref{eval:cpu-lmbench-reader-writer} shows the average CPU usage as observed by the reader (top) and writer (bottom) threads respectively at the different points during the lifetime of the dynamic benchmark.

\begin{figure}[!t]
	\centering
	\begin{subfigure}[b]{0.9\columnwidth}
		\includegraphics[scale=0.6]{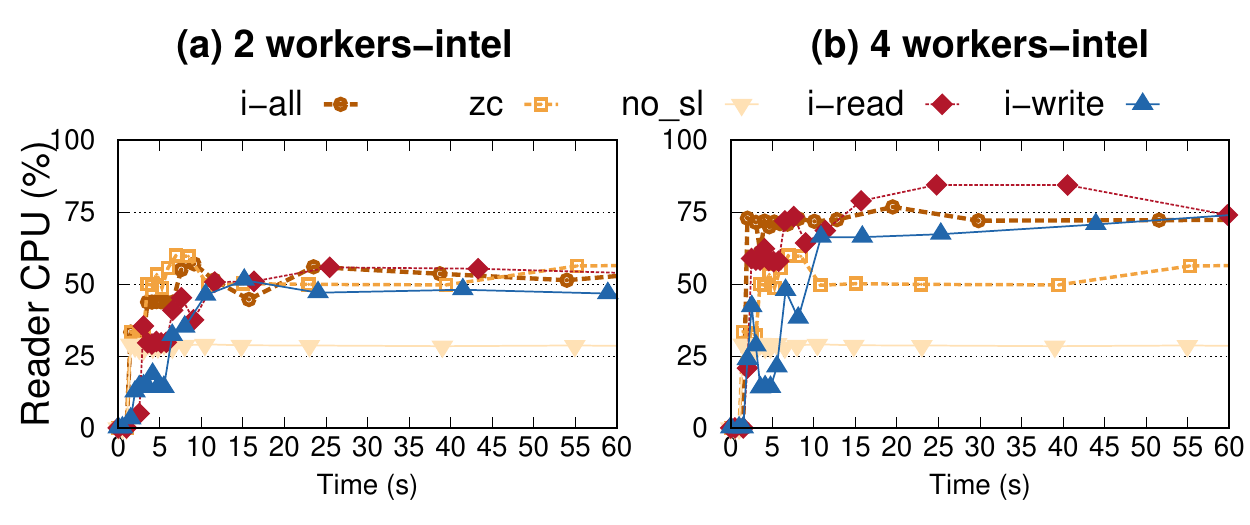}
		\label{eval:cpu-lmbench-reader}
	\end{subfigure}
	\begin{subfigure}[b]{0.9\columnwidth}
		\includegraphics[scale=0.6]{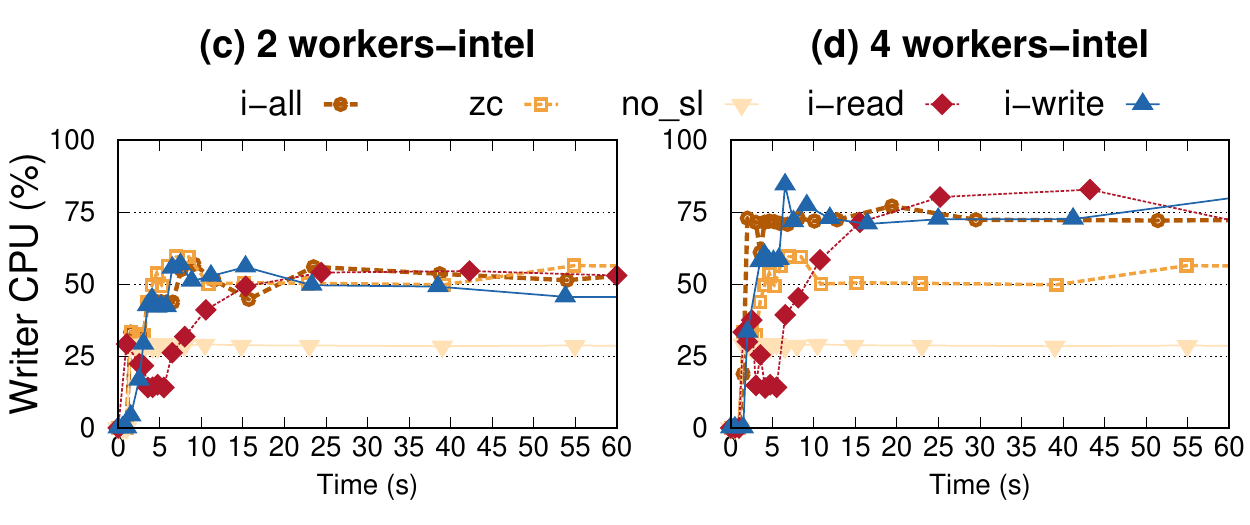}
		\label{eval:cpu-lmbench-writer}
	\end{subfigure}
	\caption{Read(top)/Write(bottom) CPU usage \% for \zc vs. 2/4 Intel switchless worker threads.}
	\label{eval:cpu-lmbench-reader-writer}
	\addfigvspace
\end{figure}

\smallskip\noindent
\textbf{Observation.}
Similarly to the throughputs, the CPU usage for the studied configurations increases with time and plateaus at a certain point. The experimental results show that on average, \zc CPU usage is about $1.8\times$ and  $1.6\times$ more when compared to reader-\iwrite-2 and writer-\iread-2 respectively, but almost equal CPU usage on average when compared to reader-\iwrite-4, writer-\iread-4, and reader/writer-\iall-2 respectively. However, reader/writer-\iall-4 use about $1.3\times$ more CPU when compared to \zc.

\smallskip\noindent
\textbf{Discussion.}
Similarly to the poor throughputs, we can easily highlight CPU waste for the misconfigurations: reader-\iwrite-4 and writer-\iread-4, as they have similar CPU usages when compared to \zc on average but much poorer performance with regards to the corresponding throughputs. \sys prevents this misconfiguration problem. For the better configurations: reader-(\iread,\iall), and writer-(\iwrite,\iall), Intel performs better than \zc but this usually comes at a higher CPU cost (especially for 4 Intel workers), which is explained by the fact that Intel's switchless mechanism maintains the maximum number of workers when there are pending switchless requests.

\observ{Poorly configured Intel switchless systems lead to a waste of CPU resources. \sys obviates the performance penalty from misconfigured Intel switchless systems, while minimising CPU waste.}


\subsection{Performance of improved \texttt{memcpy}}
\label{eval:memcpy}

\emph{Answer to Q4.}
To evaluate the performance of our improved \memcpy implementation, we ran a benchmark similar to that in \S\ref{memcpy-optimization}, which issues 100,000 \texttt{write} system calls from within the enclave, with various sizes of aligned and unaligned buffers ranging from 512\,B to 32\,kB.
We ran this benchmark in two modes: using the default \memcpy implementation of the SDK (\vanilla) and using our improved \memcpy implementation (\zcmemcpy).

As shown in \autoref{fig:eval-memcpy-buf}, our revised \texttt{memcpy} implementation achieves a speedup, in the case of larger buffers, of up to $3.6\times$ for aligned buffers and $15.1\times$ for unaligned buffers.
Noteworthy, these speedups will benefit both regular and switchless \ocalls, as well as any application using the Intel SDK's \memcpy implementation inside enclaves.

\smallskip\noindent
\textbf{Impact on inter-enclave communication.}
Recent work~\cite{haibo} presents a quantitative study on the performance of real-world serverless applications protected in SGX enclaves. A performance test of \zcmemcpy in this work showed a $7\%-15\%$ speedup for inter-enclave SSL transfers in the context of their benchmarks, which confirms the efficiency of \zcmemcpy for copy-intensive operations.

\begin{figure}[!t]
	\centering
	\includegraphics[scale=0.7]{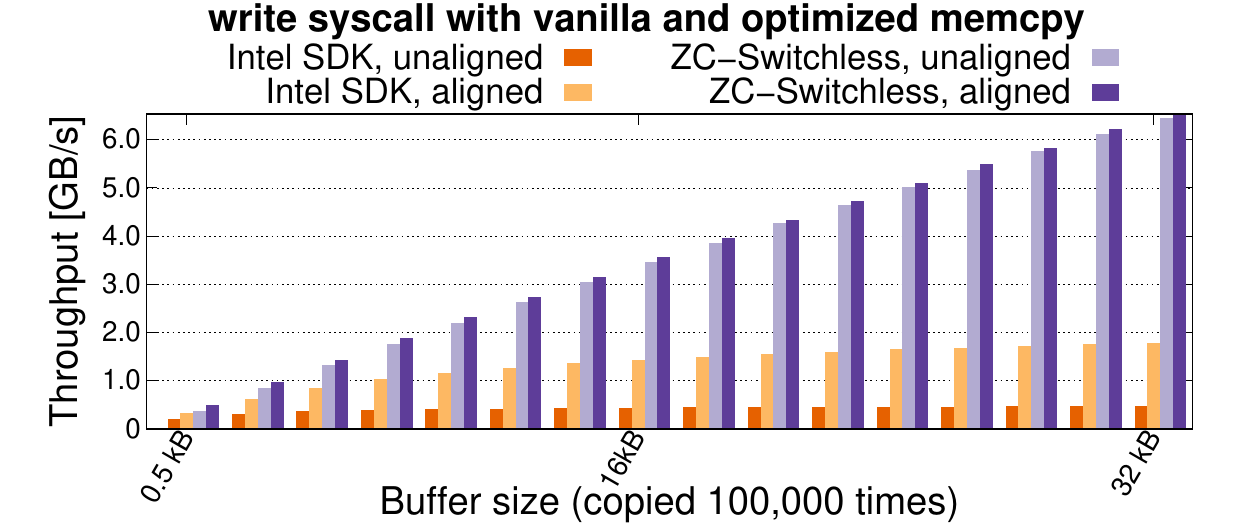}
	\caption{Throughput for \ocalls of the \texttt{write} system call to \texttt{/dev/null} (100,000 operations, average over 10 executions) for aligned and unaligned buffers.}
	\label{fig:eval-memcpy-buf}
	\addfigvspace
\end{figure}

\section{Related work}
\label{sec:rw}

We classify related work into three categories, as detailed next.

\looseness-1
\smallskip\noindent\textbf{(1) SGX benchmarking and auto-tuning tools.}
In~\cite{sgxTuner}, authors use stochastic optimisation to enhance the performance of applications hardened with Intel SGX. 
\cite{sgxOmeter} proposes a framework for benchmarking SGX-enabled CPUs, micro-code revisions, SDK versions, and extensions to mitigate enclave side-channel attacks.
These tools do not provide dynamic configuration of the switchless mechanisms.

\smallskip\noindent\textbf{(2) SGX performance improvement.}
Weichbrodt et al.~\cite{sgxperf} propose a collection of tools for high-level dynamic performance analysis of SGX enclave applications, as well as recommendations on how to improve enclave code and runtime performance, \eg by batching calls, or moving function implementations in/out of the enclave to minimise enclave transitions. \textit{Intel VTune Profiler}~\cite{vtune} permits to profile enclave applications to locate performance bottlenecks, while Dingding et al.~\cite{sgxpool} provide a framework to improve enclave creation and destruction performance.
\sys focuses on improving enclave performance efficiently via the switchless call mechanism.



\smallskip\noindent\textbf{(3) SGX transitions optimizations.}
Previous work~\cite{hotcalls,scone,eleos,sgxKernel} circumvents expensive SGX context switches by leveraging threads in and out of the enclave which communicate via shared memory, an approach also implemented by Intel~\cite{sgxdevref}.

Tian et al.~\cite{switchless} propose a switchless worker thread scheduling algorithm aimed at maximising worker efficiency so as to maximise performance speedup. \sys on the other hand, leverages a scheduling approach aimed at minimising CPU waste while improving application performance relative to a non-switchless system.



\section{Conclusion and Future Work}
\label{sec:conclusion}

In this paper, we highlight the limitations of Intel's switchless call implementation via experimental analysis. To mitigate the issues raised, we propose \sys, an efficient and configless technique to drive the execution of SGX switchless calls. 
\sys leverages an in-application scheduler that dynamically configures an optimal number of worker threads which minimises the waste of CPU resources and obviates the performance penalty from misconfigured switchless calls.
Our evaluation with a varied set of benchmarks shows that \sys provides good performance while minimising CPU waste.

We will extend \sys by integrating with profiling tools (see \S\ref{sec:rw}), to offer deployers an additional monitoring knob over SGX-enabled systems. 
Further, while the performance issues with \texttt{memcpy} were unexpected, we speculate similar issues might exist in other routines of the \texttt{tlibc}, for which a more in-depth analysis should be dedicated.

\section*{Acknowledgments}
This work was supported by the Swiss National Science Foundation under project PersiST (no. 178822) and the VEDLIoT (Very Efficient Deep Learning in IoT) European project (no. 957197).


\bibliographystyle{plain}
\bibliography{min}
\end{document}